\documentclass{article}

\usepackage{arxiv}

\usepackage[utf8]{inputenc} 
\usepackage[T1]{fontenc}    
\usepackage{hyperref}       
\usepackage{url}            
\usepackage{booktabs,longtable}       
\usepackage{siunitx}
\usepackage{amsfonts}       
\usepackage{nicefrac}       
\usepackage{microtype}      
\usepackage{lipsum}
\usepackage{xcolor}
\usepackage{float}

\usepackage{mathtools}
\usepackage{float}
\usepackage[style=nature,backend=biber,doi=false,url=false]{biblatex}
\title{Universal Nucleation Behavior of Sheared Systems}

\author{
	\href{https://orcid.org/0000-0001-8706-2383}{\includegraphics[scale=0.06]{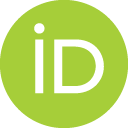}\hspace{1mm}Amrita Goswami}\\
  Department of Chemical Engineering\\
  Indian Institute of Technology Kanpur\\
  \texttt{amritag@iitk.ac.in} \\
	\And
	{Indranil Saha Dalal}\thanks{\textbf{Corresponding Author}} \\
  Department of Chemical Engineering\\
  Indian Institute of Technology Kanpur\\
  \texttt{indrasd@iitk.ac.in} \\
	\And
	\href{https://orcid.org/0000-0001-8056-2115}{\includegraphics[scale=0.06]{orcid.png}\hspace{1mm}Jayant K. Singh}\footnotemark[1] \\
	Department of Chemical Engineering\\
	Indian Institute of Technology Kanpur\\
	\texttt{jayantks@iitk.ac.in} \\
}

\begin{document}
\maketitle

\begin{abstract}
	Using molecular simulations and a modified Classical Nucleation Theory, we study the nucleation, under flow, of a variety of liquids: different water models, Lennard-Jones and hard sphere colloids. Our approach enables us to analyze a wide range of shear rates inaccessible to brute-force simulations. Our results reveal that the variation of the nucleation rate with shear is universal. A simplified version of the theory successfully captures the non-monotonic temperature dependence of the nucleation behavior, which is shown to originate from the violation of the Stokes-Einstein relation.
\end{abstract}

\keywords{rare-event, nucleation, shear, seeding, Classical Nucleation Theory}

\hypertarget{introduction}{%
\section{Introduction}\label{introduction}}

The nucleation of quiescent systems, at molecular scales, is of major interest and has been the focus of intense research
\cite{Sosso2016a}. However, in nature and in practice, static fluids are rarely involved; realistic systems almost always exist in a state of flux. The study of the effects of shear on nucleation is a burgeoning field, with far-reaching implications for industry and several branches of science. Despite investigations in this direction, the literature is rife with controversial results. Some studies indicate that the presence of shear inhibits the nucleation
rate \cite{Blaak2004, Blaak2004a}, while others assert that the
nucleation rate is enhanced by shear
\cite{Mokshin2009, Graham2009, Radu2014, Forsyth2014, Shao2015, Ruiz-Franco2018, Stroobants2020}. A non-monotonic dependence of the induction times for nucleation has also been reported in experiments \cite{Holmqvist2005, Liu2013}.


The homogeneous nucleation of the sheared Ising model \cite{Allen2008},
colloidal models \cite{Cerda2008, Lander2013}, hard spheres (HS) \cite{Richard2015, Mura2016, Richard2019}, glassy systems \cite{Mokshin2010, Mokshin2013}, a binary-alloy \cite{Peng2017}, and more
recently mW water under shear \cite{Luo2020, Goswami2020a}, has been
studied using theory and simulations. Water is a highly anomalous liquid
exhibiting several anomalies in the supercooled regime
\cite{Pettersson2016}, but efforts have not been made to distinguish the nature of the
shear-dependent nucleation behavior of water, or to generalize shared traits. However, existing literature implies that the nucleation rate for liquids, including water, is non-monotonic with shear \cite{Cerda2008, Lander2013, Richard2015, Mura2016, Richard2019, Mokshin2010, Mokshin2013, Peng2017, Luo2020, Goswami2020a}.

In this work, we generalize the phenomenon of shear-induced nucleation by revealing the underlying universality of the same. Recently we formulated a Classical Nucleation Theory (CNT), extended to explicitly incorporate shear \cite{Goswami2020a}. Here, we show the generality of this approach (henceforth referred to as 'shear-CNT'), using it to explain the effects of shear on various systems: the rigid water
models TIP4P/2005 \cite{Abascal2005}, TIP4P/Ice \cite{Abascal2005a}, the
coarse-grained mW water model \cite{Molinero2009}, the Lennard-Jones (LJ)
fluid \cite{Broughton1983}, and a HS colloid. We examine, in detail, the dual effects of temperature and shear on the nucleation rates for water and LJ fluid, explore the provenance of anomalies, and highlight the universality in the nucleation behavior.

\hypertarget{theory-and-methods}{%
\section{Theory and Methods}\label{theory-and-methods}}

The free energy of a crystal nucleus in a bulk homogeneous nucleating
system, under the effect of a simple volume-preserving shear
\(\dot{\gamma}\), is given by \cite{Mura2016}:

\begin{equation}\label{eqn:freeEnergy}
F(R) = -\frac{4}{3} \pi R^3 \frac{|\Delta \mu_{0} |}{v'} + 4 \pi R^2 \sigma_{0} \left[ 1 + \frac{7}{24} (\tau \dot{\gamma})^2\right] + \frac{1}{2} G (\tau \dot{\gamma})^2 \frac{4}{3} \pi R^3, \tag{$1$}
\end{equation}

where \(F(R)\) is the free energy of formation of a cluster of radius
\(R\), \(|\Delta \mu_{0}|\) is the chemical potential difference between
the thermodynamically stable crystal phase and the metastable liquid
phase when no shear is applied, \(\sigma_{0}\) is the surface tension or
the interfacial free energy of the nucleus at zero shear, \(v'\) is the
volume of one molecule in the crystal phase, \(G\) is the shear modulus of the nucleus, and $\tau$ is a characteristic time defined as $\tau=\frac{\eta}{G}$, where $\eta$ is the fluid viscosity. 

Homogeneous nucleation is an activated process, exhibiting a maximum in the free energy at a critical nucleus size $N^*$. The height of the free energy barrier for nucleation, corresponding to
this critical nucleus size \(N^*\), is obtained from

\begin{equation}\label{eqn:criticalF}
F(N^*) = \frac{N_0^{*} |\Delta \mu_0|}{2} \frac{ [1 + \frac{7}{24} (\tau \dot{\gamma})^2]^3 }{ \left[ 1 - \frac{v' G}{2 |\Delta \mu_0|} (\tau \dot{\gamma})^2 \right]^2 }, \tag{$2$}
\end{equation}

where $N_0^{*} = \frac{32 \pi \sigma_0^3 v'^2}{3 |\Delta \mu_0|^3}$ is the critical nucleus size at zero shear.

The steady-state nucleation rate, \(J\), can be estimated using the
following familiar CNT-based expression \cite{Goswami2020a}:

\begin{equation}\label{eqn:rate}
J = \rho_l Z f^+ e^{-\frac{F(N^*)}{k_B T}}, \tag{$3$}
\end{equation}

where the nucleation rate \(J\) is the current or flux across the free
energy barrier, in the cluster-size space and is in units of the number
of nucleation events per unit volume per unit time, \(f^+\) is the rate
of attachment of particles to the critical cluster, \(\rho_l\) is the
number density of the supercooled liquid, and \(Z\) is the Zeldovich
factor. \(Z\) captures the probability of multiple
re-crossings of the energy barrier\cite{Pan2004}.   

The expression for the shear rate-dependent attachment rate \(f^+\) is
given by \cite{Goswami2020a}:

\begin{equation}\label{eqn:diffusionCNT}
f^+ = \frac{24 D_l}{\lambda^2} (N^*)^{\frac{2}{3}} \left[ 1 + \frac{7}{24} (\tau \dot{\gamma})^2 \right], \tag{$4$}
\end{equation}

where \(D_l\) is the two-dimensional diffusion coefficient of the supercooled liquid
phase for a particular shear rate and temperature \(T\), and \(\lambda\) is the atomic `jump length', estimated to be about
one molecule diameter. 

It has been shown earlier that the diffusion coefficient varies linearly with shear rates, at a constant temperature, for the mW model \cite{Goswami2020a}:

\begin{equation}\label{eqn:diffusionLinearFit}
D_l = D_0 + c \dot{\gamma}, \tag{$5$}
\end{equation}

where \(D_0\) is the diffusion coefficient when the shear rate is zero, and \(c\) is a
fitting parameter with units of squared length. We observe that Eq.~(\ref{eqn:diffusionLinearFit}) holds true for TIP4P/2005, TIP4P/Ice, mW and LJ. We have estimated $c$ for these systems by fitting $D_l$, from out non-equilibrium molecular dynamics (NEMD) simulations, to Eq.~(\ref{eqn:diffusionLinearFit}). Such a linear behavior is predicted for a suspension of particles, which also provide the following estimate for $c$ \cite{Siqueira2017, Chandran2020}: 

\begin{equation}\label{eqn:cRelation}
c = K_c a^2 \phi, \tag{$6$}
\end{equation}

where $a$ is the particle diameter, $\phi$ is the volume fraction, and $K_c$ is a constant. We have used a value of $K_c = 0.4$, which has been successfully used for suspensions \cite{Siqueira2017} and blood \cite{Chandran2020}. In this letter, for hard-sphere colloids, we use Eq~(\ref{eqn:cRelation}) to estimate the value of $c$. $D_0$ is calculated using the Stokes-Einstein relation, given by $D_0 = (\rho_l)^{\frac{1}{3}} \frac{k_B T}{6 \eta}$, modified for hard spheres \cite{Ohtori2018}.


We note that the shear rates considered in this study are low enough to safely assume that the fluids exhibit Newtonian behavior. Further, we assume that the shear modulus $G$ of the nuclei is isotropic, which may not be strictly true for ice. However, the variations in $G$ for both hexagonal ice and amorphous ice are within the range of $3-4.5 \ GPa$ \cite{Loerting2006, Cao2018, Moreira2019}, which does not significantly impact the calculated nucleation rates \cite{Goswami2020a}.

\hypertarget{results-and-discussion}{%
\section{Results and Discussion}\label{results-and-discussion}}

The shear-CNT formulation predicts that the nucleation rate $J$ has a non-monotonic nature with respect to the shear rate, owing to competing energetic and kinetic effects. Eq.~(\ref{eqn:criticalF}) shows that the free energy barrier will rise with increasing shear rates. Eq.~(\ref{eqn:diffusionCNT}) indicates that $f^+$ will increase, due to the increase in both $D_l$ and $N^*$. The net effect is that of a maximum in $J$ (Eq.~(\ref{eqn:rate})) at some particular shear rate. 



To analyze the non-monotonicity, we introduce a min-max normalized \cite{Freedman2018} nucleation rate, \(\frac{J-J_0}{J_{max}-J_0}\), defined with respect
to \(J_0\), the nucleation rate at zero shear, and \(J_{max}\), the
highest nucleation rate observed at a particular temperature
\cite{Goswami2020a}. The optimal shear rate, \(\dot{\gamma}_{opt}\), is defined as the shear rate for which \(\frac{J-J_0}{J_{max}-J_0}\) is maximized.

We observe that, for all the systems studied in this work, parabolic
fits approximate the nucleation rate behavior with excellent agreement. A parabolic law, with respect to the dimensionless shear $\frac{\dot{\gamma}}{\dot{\gamma}_{opt}}$, of the following form can describe
the nucleation behavior at a particular temperature \(T\) and
supercooling \(\Delta T\):

\begin{equation}\label{eqn:famParabola}
\frac{J-J_0}{J_{max}-J_0} = 1 - \left( \frac{\dot{\gamma}}{\dot{\gamma}_{opt}}-1 \right)^2, \tag{$7$}
\end{equation}

The vertex of this parabola is at unity. We recover a family of parabolas with vertices at $\dot{\gamma}_{opt}$, at every temperature, if $\frac{J-J_0}{J_{max}-J_0}$ is plotted against $\dot{\gamma}$.

\begin{figure}
\centering
\includegraphics[scale=0.6]{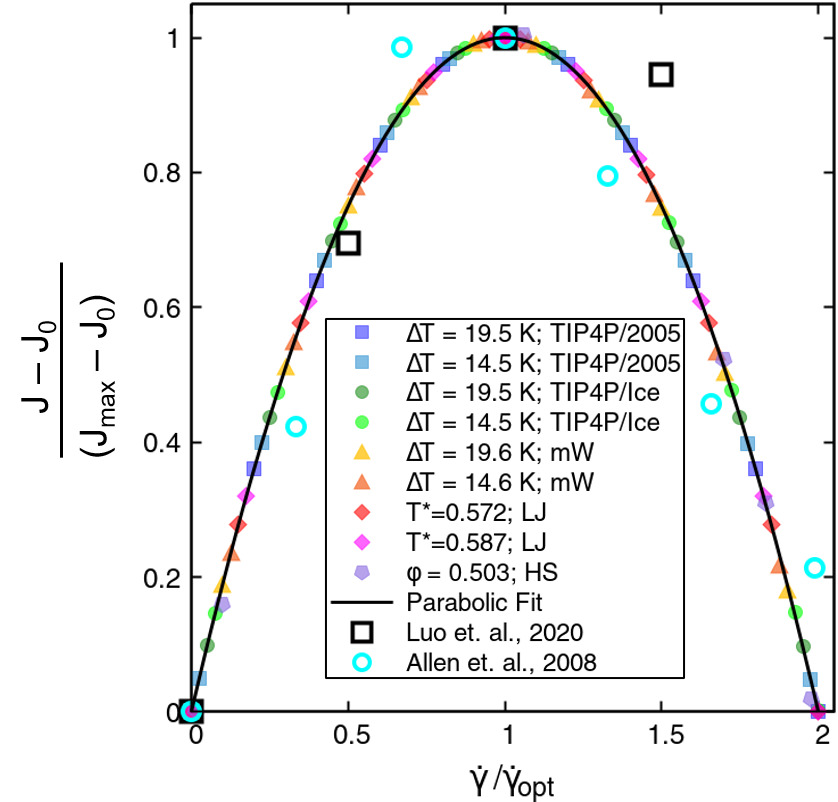}
\caption{Variation of the normalized nucleation rate with the normalized shear rate, $\dot{\gamma}/\dot{\gamma}_{opt}$ at selected metastabilities, plotted alongside the corresponding parabolic fit. Eq.~(\ref{eqn:famParabola}) has been denoted by a solid black line, and filled markers symbolize the nucleation rates calculated using shear-CNT for various systems and metastabilities. Black open squares show the data for the mW model estimated by Luo et. al. \cite{Luo2020}, for a supercooling of $67.6 \ K$, using brute-force NEMD to calculate and fit to the induction times \cite{Fitzner2015}. Open turquoise circles depict the data for a sheared two-dimensional Ising model, obtained using Forward-Flux Sampling \cite{Allen2005}, by Allen et. al. \cite{Allen2008}}
\label{fig:unitParabolaFitAll}
\end{figure}

\begin{figure*}
\centering
\includegraphics[width = \textwidth]{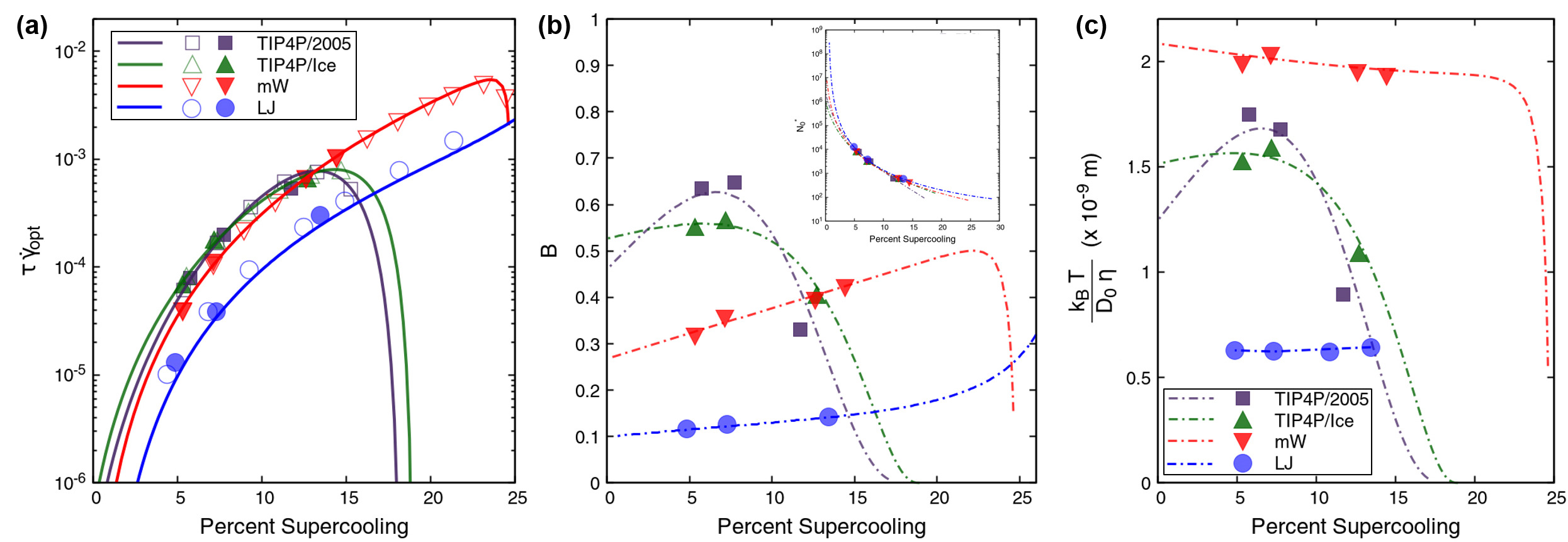}
\caption{(a) Dependence of $\tau \dot{\gamma}_{opt}$ on the percent supercooling, $\frac{T_m - T}{T_m} \times 100 \%$, for TIP4P/2005, TIP4P/Ice, mW and LJ. Filled and open markers represent values calculated using shear-CNT, with input data calculated from simulations and with approximated inputs, respectively. The solid lines denote $\tau \dot{\gamma}_{opt}$ estimated using the simplified theory, Eq.~(\ref{eqn:simpleGammaTau}). (b) Variation of $B = \left( \frac{k_B T}{D_0 \eta} \frac{c}{v'} \right)$ with the percent supercooling. The inset shows $N_0^*$ plotted against the percent supercooling. (c) Test of the Stokes-Einstein (SE) relation, according to which $\frac{k_B T}{D_0 \eta}$ should be constant. SE relation is clearly violated for the water models for the supercoolings considered. Data for which the values of $\eta$, $D_0$, $c$ and $v'$ were calculated using simulation data are denoted by filled markers, and dashed lines show values calculated using approximations.}
\label{fig:normOptShear}
\end{figure*}

Figure~\ref{fig:unitParabolaFitAll} depicts the universality in the normalized nucleation rate, generated by the superposition of available data for the water models, LJ fluid and hard spheres. These include our results, as well as those of earlier studies by other groups \cite{Luo2020, Allen2008}. We infer the existence of a single maximum nucleation rate, at any given metastability, for every system. For shear rates higher than \(\dot{\gamma}_{opt}\),
the nucleation rate decreases. Despite the complex interactions of
shear-dependent terms in Eq.~(\ref{eqn:rate}), the simple functional
form of Eq.~(\ref{eqn:famParabola}) works well for all systems.
These results indicate that this behavior is fundamental to Newtonian
fluids. 

A previous study on the mW model suggests that the shear-dependent
nucleation rates have a non-linear dependence on the temperature
\cite{Goswami2020a}. This could arise from the inclusion of several temperature-dependent
parameters in the expression for the nucleation rate
(Eq.~(\ref{eqn:rate})). Scrutiny of Eq.~(\ref{eqn:freeEnergy}), Eq.~(\ref{eqn:criticalF}) and Eq.~(\ref{eqn:diffusionCNT}) reveals the recurring dimensionless group $\tau \dot{\gamma}$. The temperature dependence of the nucleation behavior under shear is embodied by the dimensionless product, $\tau \dot{\gamma}_{opt}$, where $\dot{\gamma}_{opt}$ depends on the temperature as well as the nature of the system. However, the transcendental nature of the nucleation rate expression prevents us from directly solving an analytical expression for $\tau \dot{\gamma}_{opt}$. 

In order to further simplify the governing equations of the shear-CNT formalism and obtain a relation for $\tau \dot{\gamma}_{opt}$, we examine the order of magnitudes for the various parameters in the equations involved. For water, LJ and HS, the shape factor is $\approx 1$ for the highest shear rates considered. For $\dot{\gamma} < \frac{1}{\eta} \left( \frac{2 G |\Delta \mu_0|}{v'} \right)^{\frac{1}{2}}$, we can use a binomial expansion for the denominator in Eq.~(\ref{eqn:criticalF}). Subsequently expanding the exponential in Eq.~(\ref{eqn:rate}), we finally obtain a simplified expression for $J$:

\begin{equation}\label{eqn:simpleRate}
J = J_0 \left( 1 + \frac{c}{D_0} \right) \left[1 - \frac{N_0^* v' G}{2k_B T} (\tau \dot{\gamma})^2 \right], \tag{$8$}
\end{equation}

where $J_0$ is the nucleation rate when the shear rate is zero. We note that, although Eq.~(\ref{eqn:simpleRate}) is cubic in $\dot{\gamma}$, there exists only one positive root $\dot{\gamma}_{opt}$. This is reflected by the existence of a single maximum in the master curve for the normalized nucleation rate, shown in Fig.~\ref{fig:unitParabolaFitAll}.

As the magnitude of the dimensionless term $\frac{6 c^2 G k_B T}{N_0^* v' (D_0 \eta)^2}$ is one and two orders of magnitude lower than unity for the water models and LJ, respectively, we obtain the following relation for $\tau \dot{\gamma}_{opt}$:

\begin{equation}\label{eqn:simpleGammaTau}
\tau \dot{\gamma}_{opt} = \left( \frac{k_B T}{D_0 \eta} \frac{c}{v'} \right) \times \frac{1}{N_0^*}, \tag{$9$}
\end{equation}

where we define $B = \left( \frac{k_B T}{D_0 \eta} \frac{c}{v'} \right)$, which is a dimensionless group related to the transport properties. $N_0^*$ is dependent on the thermodynamic properties. $\eta$ and $D_0$ are approximated by power law fits. Second-order polynomials suffice to approximate the densities \cite{Espinosa2016}. Linear fits to $\sigma_0$ \cite{Espinosa2016}, $|\Delta \mu_0|$, $c$, are used to obtain the predicted values. To compare the behavior of the water models and LJ fluid, we define the percent supercooling with respect to the melting point, $T_m$, for each model.

Figure~\ref{fig:normOptShear}(a) shows the variation in $\tau \dot{\gamma}_{opt}$ with percent supercooling for the water models and LJ. The simplified Eq.~(\ref{eqn:simpleGammaTau}) performs well for the models considered (denoted by lines in Fig.~\ref{fig:normOptShear}). $\tau \dot{\gamma}_{opt}$ exhibits a single maximum for every water model. In particular, the rigid water models show nearly identical behavior. We also note that every system shows monotonic increase in the limit of $10 \%$ supercooling. However, the $\tau \dot{\gamma}_{opt}$ curve for LJ shows a qualitatively different trend compared to the water models for higher supercooling. 


Figure~\ref{fig:normOptShear}(b) depicts the dependence of the dimensionless group $B$ on the percent supercooling. The non-monotonic behavior of $B$ for the water models closely mirrors that of $\tau \dot{\gamma}_{opt}$ in Fig.~\ref{fig:normOptShear}(a). Concomitantly, we attribute the trend in $\tau \dot{\gamma}_{opt}$ for LJ to the monotonic behavior of $B$. The inset of Fig~\ref{fig:normOptShear}(b) shows a nearly universal trend of $N_0^*$ with percent supercooling.

Furthermore, our analysis shows that the origin of the divergent trends in $B$ (Fig.~\ref{fig:normOptShear}(b)) lies in the Stokes-Einstein (SE) relation. Anomalous transport properties of supercooled liquids are often characterized by SE violation \cite{Hodgdon1993, Stillinger1995, Tarjus1995, Cicerone1995, Ediger2000, Shi2013, Sengupta2013, Henritzi2015, Kawasaki2017}. According to the SE relation, the following expression holds true at all temperatures \cite{Sutherland1905, Hynes1977}:

\begin{equation}\label{eqn:seRelation}
D_0 \propto \frac{k_B T}{\eta}, \tag{$10$}
\end{equation}

which implies that, if the SE relation is valid, the term $\frac{k_B T}{D_0 \eta}$ is constant.  

Figure~\ref{fig:normOptShear}(c) depicts the variation of $\frac{k_B T}{D_0 \eta}$ with percent supercooling. The SE relation breaks down spectacularly for supercooled water \cite{Chen2006, Xu2009, Kawasaki2017, Kawasaki2019}, as shown by maxima in the $\frac{k_B T}{D_0 \eta}$ curves for TIP4P/2005, TIP4P/Ice and mW. These are directly reflected by the maxima of $B$ and $\tau \dot{\gamma}_{opt}$ for the water models. In contrast, $\frac{k_B T}{D_0 \eta}$ is relatively constant for LJ (Fig.~\ref{fig:normOptShear}(c)), which suggests that the SE relation is preserved, in the case of the LJ fluid, for the supercoolings considered in this work. We surmise that the temperature dependence of the nucleation behavior is strongly linked to the violation or preservation of the SE relation, and thus depends solely on the behavior of flow properties. The decoupling of $D_0$ and $\eta$, typified by the SE violation, is thought to originate from spatial heterogeneities in the dynamics of strongly supercooled glass-forming liquids \cite{Hodgdon1993, Stillinger1994, Ediger2000, Lombardo2006, Sengupta2013, Kawasaki2017, Kawasaki2019}.

\hypertarget{conclusions}{%
\section{Conclusions}\label{conclusions}}

In conclusion, we have reported the effects of shear on the nucleation
rates at different temperatures, for the TIP4P/2005, TIP4P/Ice, mW water
models, LJ fluid and HS colloids. Nucleation events at low and moderate
supercoolings are notoriously difficult to simulate, and such extensive
calculations are virtually intractable using brute-force molecular dynamics. By employing the shear-CNT formalism, based on modified CNT equations, we
were able to obtain nucleation rate curves for several metastable conditions.

In accordance with previous simulation results for colloids, glassy
systems, the Ising model, and mW water
\cite{Allen2008, Richard2015, Mura2016, Luo2020, Goswami2020a}, we confirmed that
the nucleation rate curves exhibit non-monotonic behavior with shear,
at a particular supercooling. We generated a "universal" master curve for the normalized nucleation rate \(\frac{J-J_0}{J_{max}-J_0}\) with $\dot{\gamma}/\dot{\gamma}_{opt}$.
Despite the complicated dependence of shear in the transcendental
equation for the nucleation rate, parabolic fits yield excellent
agreement to the nucleation rate curves and are valid for every system
considered in this work. We infer that the existence of a maximum in the
nucleation rate with shear is a universal property of systems that obey CNT.

We systematically investigated the temperature dependence of the nucleation
rate curves for TIP4P/2005, TIP4P/Ice, mW and LJ by examining the behavior of the dimensionless group $\tau \dot{\gamma}_{opt}$. To this end, we derived a simplified theory describing the governing equations of shear-CNT. An approximate relation for $\tau \dot{\gamma}_{opt}$ was obtained, expressed as a product of two dimensionless groups: $B$, which is related to transport properties, and the thermodynamic quantity $1/N_0^*$. The analysis reveals that the behavior of $\tau \dot{\gamma}_{opt}$ is solely determined by the nature of $B$. The anomalous temperature dependence of the nucleation behavior of water originates from the SE violation. We discovered that universal behavior is recovered for $N_0^*$, for every system.


Thus, we have uncovered underlying commonalities and determined the origin of anomalies in the nucleation behavior for several supercooled molecular systems under shear. Our results
provide insight into the previously unexplored, intriguingly complex
interplay of temperature and shear, affecting the nucleation rate.

\hypertarget{acknowledgements}{%
\section{Acknowledgements}\label{acknowledgements}}

This work was supported by the Science and Engineering Research Board
(sanction number STR/2019/000090 and CRG/2019/001325). Computational
resources were provided by the HPC cluster of the Computer Center (CC),
Indian Institute of Technology Kanpur.

\printbibliography


\end{document}